\documentclass[aps,prl,superscriptaddress,twocolumn,showpacs,showkeys]{revtex4-2}
\usepackage{graphicx}
\usepackage{amsmath, amsxtra, amssymb, mathtools, isomath, nccmath, xspace, siunitx}
\usepackage{color}
\usepackage[utf8]{inputenc}
\usepackage{siunitx}  
\usepackage[british]{babel}
\usepackage{hyperref}

\usepackage{sidecap}
\usepackage{natbib}
\usepackage{dsfont}
\usepackage{xcolor}
\usepackage{pifont}

\newcommand{\ket}[1]{\ensuremath{\lvert #1 \rangle}\xspace}%
\newcommand{\abs}[1]{\ensuremath{\lvert #1 \rvert}\xspace}%

\sidecaptionvpos{figure}{h}
\long\def\symbolfootnote[#1]#2{\begingroup%
\def\thefootnote{\fnsymbol{footnote}}\footnotetext[#1]{#2}\endgroup}
\newcommand{\Aref}[1]{\hyperref[#1]{A}}
\newcommand{\Bref}[1]{\hyperref[#1]{B}}
\newcommand{\Cref}[1]{\hyperref[#1]{C}}
\newcommand{\Dref}[1]{\hyperref[#1]{D}}

\makeatletter
\@addtoreset{equation}{myequation}
\makeatother

\begin{document}
\setlength{\belowdisplayskip}{9pt}
\setlength{\abovedisplayskip}{9pt}
\setlength{\belowdisplayskip}{9pt}
\setlength{\abovedisplayskip}{9pt}

\title{\bf{Realizing distance-selective interactions in a Rydberg-dressed atom array}}



\author{Simon Hollerith}
\email[]{Simon.Hollerith@mpq.mpg.de}
\affiliation{Max-Planck-Institut f\"{u}r Quantenoptik, 85748 Garching, Germany}
\affiliation{Munich Center for Quantum Science and Technology (MCQST), 80799 Munich, Germany}

\author{Kritsana Srakaew}
\affiliation{Max-Planck-Institut f\"{u}r Quantenoptik, 85748 Garching, Germany}
\affiliation{Munich Center for Quantum Science and Technology (MCQST), 80799 Munich, Germany}

\author{David Wei}
\affiliation{Max-Planck-Institut f\"{u}r Quantenoptik, 85748 Garching, Germany}
\affiliation{Munich Center for Quantum Science and Technology (MCQST), 80799 Munich, Germany}

\author{Antonio Rubio-Abadal}
\thanks{present address: ICFO – Institut de Ciencies Fotoniques, The Barcelona Institute of Science and Technology, 08860 Castelldefels (Barcelona), Spain}
\affiliation{Max-Planck-Institut f\"{u}r Quantenoptik, 85748 Garching, Germany}
\affiliation{Munich Center for Quantum Science and Technology (MCQST), 80799 Munich, Germany}

\author{Daniel Adler}
\affiliation{Max-Planck-Institut f\"{u}r Quantenoptik, 85748 Garching, Germany}
\affiliation{Munich Center for Quantum Science and Technology (MCQST), 80799 Munich, Germany}
\affiliation{Fakult\"{a}t f\"{u}r Physik, Ludwig-Maximilians-Universit\"{a}t M\"{u}nchen, 80799 M\"{u}nchen, Germany}%

\author{Pascal Weckesser}
\affiliation{Max-Planck-Institut f\"{u}r Quantenoptik, 85748 Garching, Germany}
\affiliation{Munich Center for Quantum Science and Technology (MCQST), 80799 Munich, Germany}

\author{Andreas Kruckenhauser}
\affiliation{Institute for Quantum Optics and Quantum Information of the Austrian Academy of Sciencies, Innsbruck, Austria}
\affiliation{Center of Quantum Physics, University of Innsbruck, Innsbruck, Austria}

\author{Valentin Walther}
\affiliation{ITAMP, Harvard-Smithsonian Center for Astrophysics, Cambridge, Massachusetts 02138, USA}
\affiliation{Department of Physics, Harvard University, Cambridge, Massachusetts 02138, USA}

\author{Rick van Bijnen}
\affiliation{Institute for Quantum Optics and Quantum Information of the Austrian Academy of Sciencies, Innsbruck, Austria}
\affiliation{Center of Quantum Physics, University of Innsbruck, Innsbruck, Austria}

\author{Jun Rui}
\affiliation{Max-Planck-Institut f\"{u}r Quantenoptik, 85748 Garching, Germany}
\affiliation{Munich Center for Quantum Science and Technology (MCQST), 80799 Munich, Germany}
\affiliation{Hefei National Laboratory for Physical Sciences at Microscale, University of Science and Technology of China, Hefei, Anhui 230026, People’s Republic of China}

\author{Christian Gross}%
\affiliation{Max-Planck-Institut f\"{u}r Quantenoptik, 85748 Garching, Germany}
\affiliation{Munich Center for Quantum Science and Technology (MCQST), 80799 Munich, Germany}
\affiliation{Physikalisches Institut, Eberhard Karls Universit\"{a}t T\"{u}bingen, 72076 T\"{u}bingen, Germany}

\author{Immanuel Bloch}%
\affiliation{Max-Planck-Institut f\"{u}r Quantenoptik, 85748 Garching, Germany}
\affiliation{Munich Center for Quantum Science and Technology (MCQST), 80799 Munich, Germany}
\affiliation{Fakult\"{a}t f\"{u}r Physik, Ludwig-Maximilians-Universit\"{a}t M\"{u}nchen, 80799 M\"{u}nchen, Germany}%

\author{Johannes Zeiher}
\affiliation{Max-Planck-Institut f\"{u}r Quantenoptik, 85748 Garching, Germany}
\affiliation{Munich Center for Quantum Science and Technology (MCQST), 80799 Munich, Germany}

\date{\today}


\begin{abstract}
Measurement-based quantum computing relies on the rapid creation of large-scale entanglement in a register of stable qubits.
Atomic arrays are well suited to store quantum information, and entanglement can be created using highly-excited Rydberg states.
Typically, isolating pairs during gate operation is difficult because Rydberg interactions feature long tails at large distances.
Here, we engineer distance-selective interactions that are strongly peaked in distance through off-resonant laser coupling of molecular potentials between Rydberg atom pairs.
Employing quantum gas microscopy, we verify the dressed interactions by observing correlated phase evolution using many-body Ramsey interferometry.
We identify atom loss and coupling to continuum modes as a limitation of our present scheme and outline paths to mitigate these effects, paving the way towards the creation of large-scale entanglement.
\end{abstract}
\maketitle

The one-way or measurement-based quantum computer~\cite{one_way_quantum_2001} has been suggested as an alternative to usual gate-based digital quantum computers.
Contrary to the latter approach, the entanglement required for a calculation is created upfront by creating a highly-entangled cluster state~\cite{Cluster_state_Rey_2019}, and the subsequent circuit is imprinted through controlled local measurements and subsequent feedback.
Realizing such a scheme requires a single massively parallel entangling operation, which relies on controllable interactions~\cite{vanBijnen2015,SchleyerSmith2021} between all neighboring qubits in the register.
The neutral-atom quantum computing platform is naturally amenable to parallel gate operation, as demonstrated in one dimension using collisional gates~\cite{Cluster_State_munich_2003} or Rydberg atoms~\cite{Lukin_Rydgate2019}.
The dipolar nature of Rydberg interactions provides the toolbox for angular interaction control~\cite{de_leseleuc_observation_2019,Zeiher2016a}.
However, their long-range character makes it challenging to isolate atom pairs at a fixed distance for gate operations. 
This holds true for Rydberg dressing schemes where interactions are optically admixed to the ground state~\cite{Jau2016}.
Rydberg dressing has been demonstrated to create Bell pairs in optical tweezers~\cite{Jau2016}, to engineer long-range interacting Ising Hamiltonians~\cite{Coherent_Zeiher, Dressing_Schleyer-Smith} or to study the competition between dressed interactions and motion in an optical lattice~\cite{Waseem_itinerant_dressing}.
A variety of further theoretical proposals to realize spin models~\cite{Glaetzle2014,Glaetzle2017,Dressing_Squeezing2014} or extended Hubbard models~\cite{Dressing_supersolid1,Knap_dressing} rely on enhanced interaction control.

\begin{figure*}[htb]
  \includegraphics[width=1.0\textwidth]{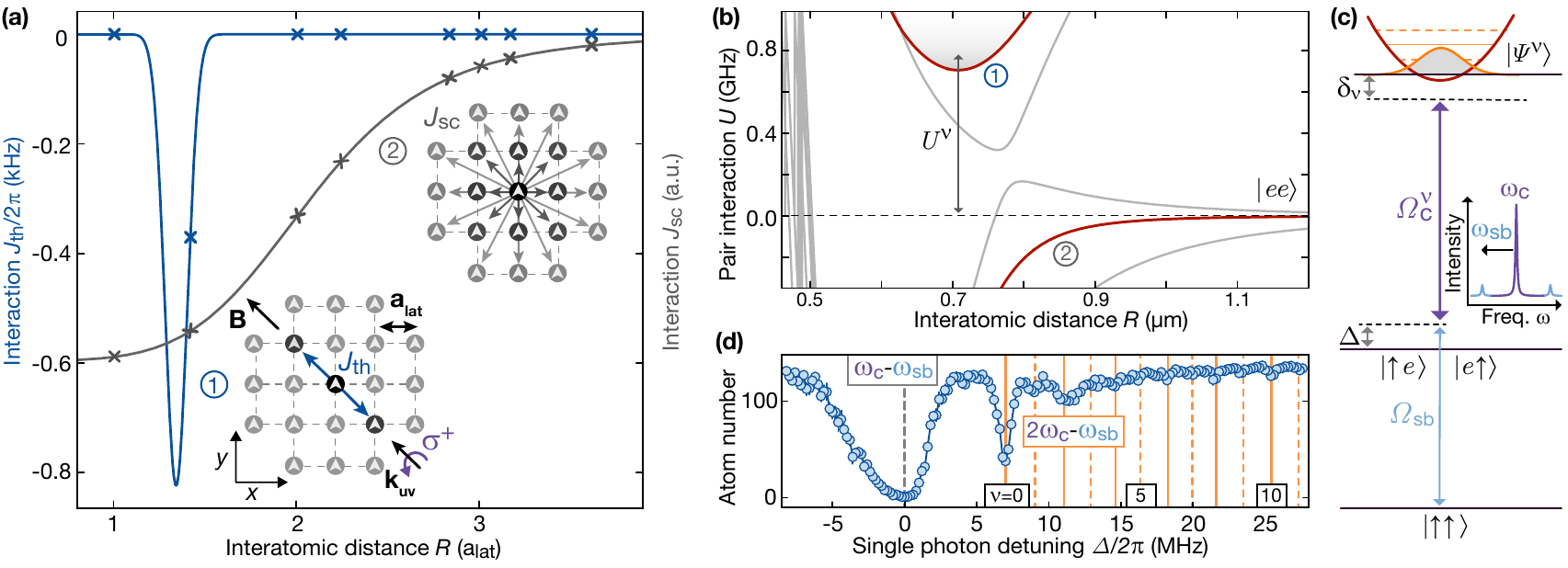}%
  \caption{\label{fig:1}
Two-color Rydberg macrodimer dressing
\textbf{(a)}
Utilizing macrodimer potentials for Rydberg dressing provides strongly localized interactions (blue), which are in stark contrast to typical soft-core interactions $J_{\textrm{sc}}$ obtained by coupling to asymptotic interaction curves (gray). 
Crosses denote the distances present in the array.
The spins are arranged in a two-dimensional square array with a spacing $a_{\textrm{lat}}$ and are illuminated by the UV laser with wavevector $\mathbf{k_{\textrm{uv}}}$ oriented along the diagonal direction of the array and parallel to the magnetic field $\mathbf{B}$.
At an interatomic distance $R = \sqrt{2}a_{\textrm{lat}}$ and an orientation $\mathbf{R}\parallel\mathbf{B}$ where the molecular Rabi couplings feature a narrow maximum, we expect to achieve a spin coupling $J_{\textrm{th}}=2\pi\times 370(40)\,\si{\Hz}$. 
\textbf{(b)} At large distances, Rydberg interaction potentials are described by van-der-Waals interactions (gray marker). 
At smaller distances, one finds macrodimer binding potentials energetically shifted by $U^\nu$ from the asymptotic state $|ee\rangle$ (blue marker). 
\textbf{(c)} We perform a two-photon excitation scheme from the ground state $\ket{\uparrow \uparrow }$ via intermediate states $\ket{\uparrow\! e }$,$\ket{e\!\uparrow }$ detuned by $\Delta$ to molecular states $|\Psi^\nu\rangle$ using Rabi couplings $\Omega_{\textrm{sb}}$ and $\Omega^\nu_{\textrm{C}}$.
 In our dressing sequence, we work at finite two-photon detunings $\delta_\nu$ to the molecular states.
 The two excitation fields are generated by modulating sidebands at frequencies $\omega_{\textrm{C}} \pm \omega_{\textrm{sb}}$ on our UV frequency $\omega_{\textrm{C}}$.
\textbf{(d)} Performing atom-loss spectroscopy, we find the vibrational spectrum slightly blue-detuned from the single-photon Rydberg transition coupled by the red sideband (here for $\omega_{\textrm{sb}} = 2\pi\times 723\,\textrm{MHz}$). 
}
\end{figure*}

Here, we demonstrate novel Rydberg-dressed interactions by coupling to bound Rydberg atom pairs, so-called Rydberg macrodimers~\cite{Boisseau2002,Overstreet2009,Sassmannshausen2016,Macrodimers_singleatoms_2019}. 
In contrast to the soft-core potentials generated in standard dressing schemes~\cite{Jau2016,Zeiher2016a}, the resulting interactions are strongly selective in distance, see Fig.~\ref{fig:1}\,(a).
We verify the presence of the dressed interactions in our two-dimensional optical lattice using many-body Ramsey interferometry~\cite{Zeiher2016a}. 
In agreement with our calculations, we observe the build-up of two-spin and three-spin correlations at the fixed chosen distance.
Finally, we identify how off-resonant scattering and photodissociation into unbound continuum states affect our dressing scheme, and discuss methods to mitigate the associated decoherence effects.

Traditionally, experiments using Rydberg atoms operate at large interatomic distances where their interaction potentials are well described by their asymptotic van-der-Waals character~\cite{Weber2017}.
In the non-perturbative regime at closer distances and large interaction energies, crossings of pair potentials naturally occur.
Avoided crossings then give rise to macrodimer binding potentials~\cite{Macrodimers_singleatoms_2019}, see Fig.~\ref{fig:1}\,(b). 
Dressing to a vibrational series of these macrodimers leads to a fundamentally different interaction profile, which peaks at the distance matching the minimum of the binding potential, see Fig.~\ref{fig:1}\,(a). 
At the same time, long-distance tails are absent because the coupling to asymptotic pair potentials is negligibly small~\cite{vanBijnen2015,Barbier2021}.
The width of the narrow interaction peaks is typically limited by the width of the ground state wave packet in the optical trap, which requires exquisite control over the motional states.
Furthermore, the dressed interactions depend critically on the orientation of the molecular states relative to applied fields and light polarizations~\cite{Structure_Tomography_Hollerith}.


Our experiments started with a two-dimensional square atom array of about two hundred $^{87}\mathrm{Rb}$ atoms in the electronic ground state $\ket{\uparrow } = |5S_{1/2},F=2,m_F=-2\rangle$ with a lattice spacing $a_{\textrm{lat}}=532\,$nm and a filling of $94(1)\%$~\cite{weitenberg11}. 
The magnetic field $\mathbf{B}$ with absolute value $\abs{\mathbf{B}} = 0.5\,$G and the wavevector $\mathbf{k_{\textrm{uv}}}$ of the excitation laser at an ultraviolet (UV) wavelength $\lambda = 298\,$nm were pointing along the lattice diagonal direction. 
The UV laser was $\sigma^+ -$polarized along the magnetic field.
The vibrational modes $\nu$ in the chosen macrodimer potential are energetically shifted by $U^\nu$ relative to the asymptotic pair state $ |ee\rangle \equiv |36P_{1/2}36P_{1/2}\rangle$.
We performed a two-photon and two-color excitation by modulating sidebands on our UV carrier frequency $\omega_{\mathrm{C}}$, see Fig.~\ref{fig:1}\,(c)~\cite{Barbier2021}.
The modulation frequency $\omega_{\mathrm{sb}}$ was sligthly below the interaction energy $U^0$ of the lowest vibrational state.   
Molecular states can then be excited by one sideband and one carrier photon, while other combinations remain off-resonant and do not contribute.
Keeping $\omega_{\mathrm{sb}}$ fixed and tuning the overall laser frequency, the vibrational modes are resonant at detunings $\frac{\Delta}{2\pi} = \frac{1}{2}(U^\nu-\frac{\omega_{\mathrm{sb}}}{2\pi})$ relative to the single-photon resonance between $\ket{\uparrow}$ and $|e \rangle $ driven by the red sideband, see Fig.~\ref{fig:1}\,(d).
The observed suppression of excitation rates for higher vibrational modes is explained by increasing detunings $\Delta$ and smaller Franck-Condon integrals with the ground state wave packet.
The two-color excitation scheme enables independent tunability of the intermediate-state detuning, the admixed scattering rates and the contributing light shifts~\cite{Barbier2021}. 
Furthermore, it allowed us to strongly increase the coupling rates into the molecular states.

The molecular bond length $R_{\nu} = 712(5)\,\textrm{nm} \approx \sqrt{2}a_{\textrm{lat}}$ restricts the coupling to molecular states oriented along the two lattice diagonals.
For the chosen configuration of light polarization and magnetic field, two-photon Rabi couplings $\Omega_\nu$ between an atom pair $\ket{\uparrow\uparrow }$ into molecular states $|\Psi^{\nu}\rangle$ reach a strong maximum for $\mathbf{R}_\parallel = (+1,-1)\,a_{\textrm{lat}}$ parallel to $\mathbf{B}$, while coupling rates at orthogonal orientation are suppressed~\cite{Structure_Tomography_Hollerith}.
This results in strong spin interactions $J_\parallel \equiv J$, while interactions $J_\perp \approx 0.06J$ along the orthogonal lattice diagonal direction are negligible on the timescale of our experiments.
The interactions $J$ arise at finite two-photon detunings $\delta_\nu$, where the molecular states are only virtually populated in a four-photon process~\cite{vanBijnen2015} and the energy of a spin-up pair $\ket{\uparrow  \uparrow }$ is reduced through $J \approx \sum_{\nu} \Omega^2_\nu/(4 \delta_\nu)$ that was dominated by the lowest vibrational mode. 
Our single-photon Rabi frequency between $\ket{\uparrow }$ and $|e\rangle$ was calibrated to be $\Omega = 2\pi\times 2.83(5)\,$MHz, our two-photon Rabi frequency is typically $\Omega_\nu \approx 2\pi\times \, 50\textrm{kHz}$.

\begin{figure}[htp]
  \includegraphics[width = 1 \columnwidth]{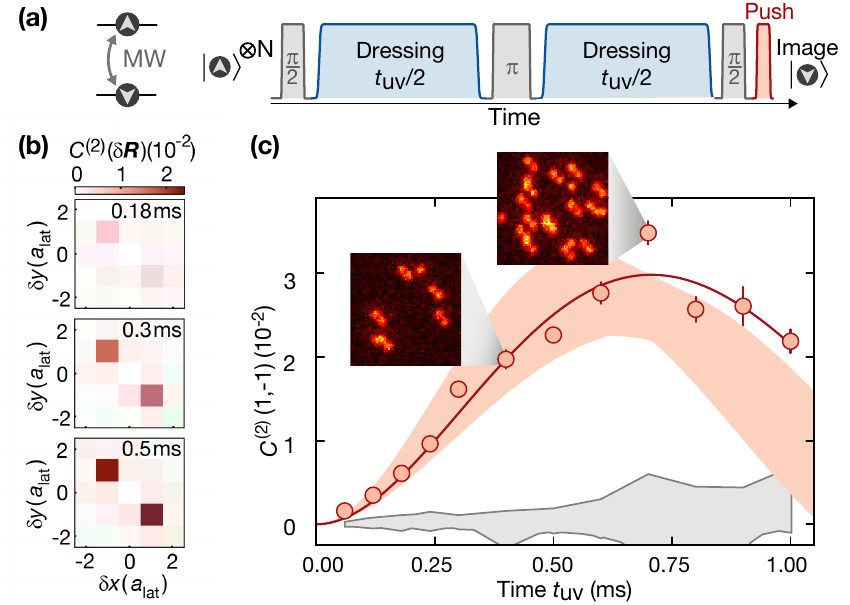}%
  \caption{\label{fig:2}
Two-spin correlations. \textbf{(a)} We encode our spin states in two hyperfine ground states coupled by a microwave field. 
Spin interactions are probed using many-body Ramsey interferometry. 
\textbf{(b)} We evaluate spin-spin correlations $C^{(2)}_\mathbf{R}$ for increasing interaction time and find a strong signal at a distance $ \mathbf{R} = (1,-1)\,a_{\textrm{lat}}$ matching the strongly coupled lattice diagonal. 
The value at the origin was excluded. 
\textbf{(c)} The observed spin dynamics  $C^{(2)}_{+1,-1}(t)$ originates from correlated spin flips during the Ramsey sequence, as shown in exemplary images from our quantum gas microscope. 
Error bars in the correlation signal were calculated using a bootstrap algorithm (delete-1 jackknife).  
We fit the observed spin dynamics to a master equation and obtain $J=2\pi\times 318(20)\,$Hz and $\Gamma^{\textrm{fit}}_{|\rightarrow\rangle} = 0.46(5)\,\si{\milli \second}^{-1}$ (solid line). 
The red shaded area corresponds to the calculated dynamics using the same model with the calculated spin coupling $J_{\textrm{th}}$ and the experimentally calibrated atom loss $\Gamma_{|\rightarrow\rangle}^{\textrm{ex}}=0.6(1)\,\si{\milli\second}^{-1}$. 
Here, uncertainties originate from $J_{\textrm{th}}$ and $\Gamma_{|\rightarrow\rangle}^{\textrm{ex}}$.
The gray shaded region represents measured two-spin correlations $C^{(2)}_\mathbf{R}$ at other distances.}
\end{figure}

\begin{figure}[htb]
  \includegraphics[width = 1 \columnwidth]{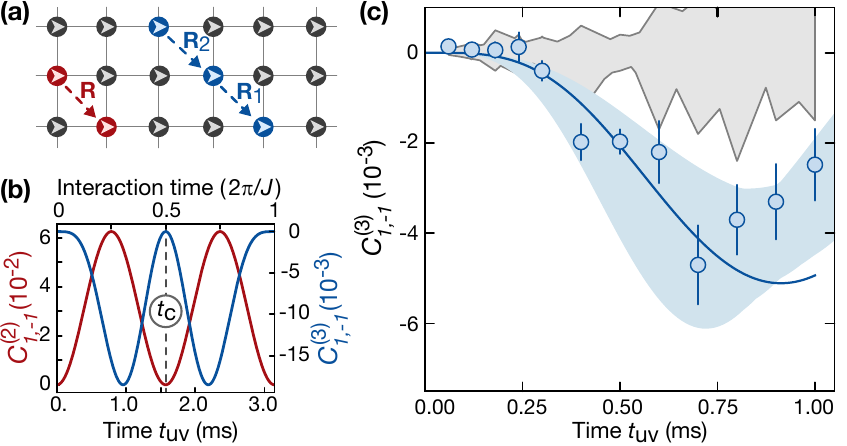}
  \caption{\label{fig:3} Higher-order correlations. \textbf{(a)} In addition to $C^{(2)}_{1,-1}(t)$ (red), we expect to also observe connected three-spin correlations  $C^{(3)}_{\mathbf{R_{1}}\mathbf{R_{2}}}$ (blue) for distance vectors $\mathbf{R}_{1} = \mathbf{R}_{2} = (1,-1)\,a_{\textrm{lat}}$ in our spin system.
\textbf{(b)} A calculation for both correlators without dissipation reveals that $C^{(3)}_{1,-1}(t)$ is expected to appear with a delay relative to $C^{(2)}_{1,-1}(t)$.
At a later time $t_{\textrm{C}}$, all coupled spins evolve into a cluster state.  
\textbf{(c)} Observed correlation dynamics of $C^{(3)}_{1,-1}(t)$. 
The solid line represents a calculation using the model parameters obtained by fitting $C^{(2)}_{1,-1}(t)$.
The dynamics is in qualitative agreement with the calculation without dissipation but the amplitude of the signal is damped. 
The blue-shaded region represents the theoretical expectation. 
The gray shaded region represents the background at other distances $\mathbf{R_{2}}$ while $\mathbf{R_{1}}=(1,-1)\,a_{\textrm{lat}}$. 
Error bars in the correlation signal are calculated using a bootstrap algorithm (delete-1 jackknife).}
\end{figure}

In a first experiment, we characterized the induced distance-selective interaction potential.
To this end, we tuned our laser to a fixed intermediate-state detuning $\Delta/2\pi = 3.58\,$MHz between the single-photon Rydberg resonance and the lowest vibrational resonance. 
The two-photon detunings are given by $\delta_\nu = \delta_0-\nu \hbar \omega_{\nu}$ relative to the vibrational series, where $\omega_{\nu}$ is the vibrational spacing and $\delta_0$ is the two-photon detuning relative to the lowest vibrational resonance $\nu = 0$. 
We realized a spin-1/2 system by including the hyperfine ground state $\ket{\downarrow } = |5S_{1/2},F=1,m_F=-1\rangle$, that was coupled to $\ket{\uparrow }$ by a microwave (MW) field but remained uncoupled to the molecular states. 
Neglecting irrelevant terms linear in the spin operators, the resulting spin lattice is thus described by the Ising Hamiltonian
\begin{equation}\label{Ising_H}
\hat{H} = \hbar\sum_{\mathbfit{i}\neq\mathbfit{j}}\frac{J_{\mathbfit{ij}}}{2}\hat{S}^z_\mathbfit{i}\hat{S}^z_\mathbfit{j},
\end{equation}
where interactions $J_{\mathbfit{ij}} = J\delta_{\mathbfit{i-j,\mathbf{R}_\parallel}}$ are restricted to the coupled lattice diagonal and $\hat{S}^z_{\mathbfit{i}}(\hat{S}^z_{\mathbfit{j}})$ are the $z-$components of the spin operators at lattice sites $\mathbfit{i}(\mathbfit{j})$.
We studied the evolution of our atom array under Eq.~\eqref{Ising_H} by performing Ramsey interferometry, see Fig.~\ref{fig:2}\,(a). 
After initializing all atoms in $\ket{ \uparrow }$, a global $\pi/2-$pulse prepared the state $\ket{ \rightarrow }^{\otimes N}$, with $\ket{ \rightarrow } = \frac{1}{\sqrt{2}}(\ket{\downarrow} - i\ket{\uparrow})$ and $N$ the total atom number in the system.
Subsequently, we applied two UV dressing pulses with duration $t_{\textrm{uv}}/2$, interrupted by a $\pi-$rotation (spin echo) in order to cancel phases originating from single-atom shifts poportional to $\hat{S}^z_\mathbfit{i}$~\cite{Zeiher2016a}.
During the evolution, coupled spin pairs accumulate phases $\varphi (t_{\textrm{uv}}) =\pm J t_{\textrm{uv}}$.
We then closed the interferometer sequence using a final $\pi/2-$rotation, removed all atoms in the spin state $\ket{\uparrow }$ and measured the remaining atoms populating the state $\ket{\downarrow }$ using the single-site resolution of our quantum gas microscope~\cite{weitenberg11}.
In this projective measurement, we observe correlated spin flips using spatially averaged connected two-point correlators $C^{(2)}_\mathbf{R} = (\langle\hat{S}^z_{\mathbf{R}^\prime}\hat{S}^z_{\mathbf{R}^\prime+\mathbf{R}}\rangle- \langle\hat{S}^z_{\mathbf{R}^\prime}\rangle\langle\hat{S}^z_{\mathbf{R}^\prime+ \mathbf{R}}\rangle)_{\mathbf{R}^\prime}$, where $(.)_{\mathbf{R}^\prime}$ denotes spatial averaging over all positions $\mathbf{R}^\prime$ in the lattice. 
\begin{figure}[htb]
  \includegraphics[width = 1 \columnwidth]{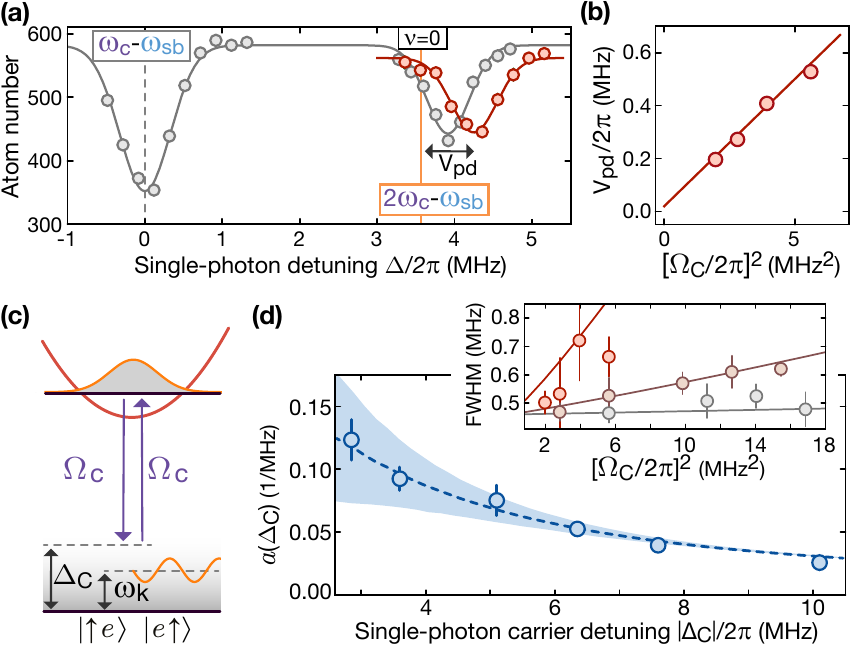}%
 \caption{\label{fig:4} Photodissociation into continuum modes. \textbf{(a)} Performing spectroscopy of the lowest vibrational line reveals a shift $V_{\textrm{pd}}$ from the expected line position (orange), which increases with laser power (gray to red), here for $\omega_{\textrm{sb}} = 2\pi\times 728\,\si{\mega\hertz}$. 
\textbf{(b)} This linear light shift $V_{\textrm{pd}}/(2\pi) = a(\Delta_C) \left[\Omega_C/(2\pi)\right]^2$ agrees very well with the theoretical model (solid line) \textbf{(c)} The calculation assumes a coupling $\Omega_{\textrm{C}}$ into singly-excited pair states occupying motional continuum states by the carrier field. 
\textbf{(d)} By varying $\omega_{\textrm{sb}}$, we measure the dependency of $a(\Delta_C)$ on the carrier detuning $\Delta_\textrm{C}$ between the molecular state and the intermediate state and find agreement with the calculation (dashed blue line). 
The blue shaded region indicates the varying contributions from different partial waves, which contribute to a broadening of the resonance, as shown in the inset for $\Delta_{\textrm{C}}/2\pi = -3.6\,$MHz, $-6.35\,$MHz, $-10.1\,$MHz (red to gray). 
Here, solid lines represent the theoretical expectation and error bars on the data points indicate the $1\sigma-67\%$ confidence interval of fitted resonance profiles.}
\end{figure}
As expected for our selective interactions, we find that correlations are restricted to distances $\mathbf{R}_\parallel$, see Fig.~\ref{fig:2}\,(b). 
After an initial quadratic increase, correlations $C^{(2)}_{1,-1}(t)$ reach a maximum at $t_{\textrm{uv}}=0.7\,$ms, which is consistent with a simulation assuming coherent spin dynamics.
At later times $t_{\textrm{uv}}$, the signal is damped due to atom loss.
Fitting a model including dissipation to the spin dynamics yields a spin interaction of $J = 2\pi \times 318(20)\,$Hz, close to the calculated value of $J_{\textrm{th}} = 2\pi\times 370(40)\,$Hz (see Fig.~\ref{fig:2}\,(c)).

A striking signature of our distance-selective spin interactions is the absence of a long-range tail.
As a result, a coherent dephasing of the many-body dynamics can be avoided and one expects to observe revivals at $t_\textrm{R} = 2\pi/J$ in the bulk of the system.
The realization of such a clean Ising Hamiltonian is particularly interesting because the coupled spins are expected to evolve into a highly entangled cluster state at times $t_\textrm{C} = t_\textrm{R}/2$~\cite{Cluster_State_munich_2003}.
Although two-spin correlations at uncoupled sites vanish~{\cite{richerme_non-local_2014}, one expects the formation of multi-spin correlations.
At $t_\textrm{C}$ where $C^{(2)}_{1,-1}(t_\textrm{C})=0$, the system still features correlations on a global scale~{\cite{n_body_LiebRobinson_Gorshkov}.
Here, we studied the emergence of higher-order correlations through the spatially averaged connected three-spin correlator $C^{(3)}_{\mathbf{R_{1}}\mathbf{R_{2}}}$~\ref{fig:S1}. 
We focus on $C^{(3)}_{1,-1}$ at distances $\mathbf{R_{1}} = \mathbf{R_{2}} = (1,-1)\,a_{\textrm{lat}}$ where both pairs are coupled by $J$, as illustrated in Fig.~\ref{fig:3}\,(a).
In a dissipationless system, a finite $C^{(3)}_{1,-1}$ value can be directly linked to 3-partite entanglement~{\cite{n_body_LiebRobinson_Gorshkov}.
Calculations in a bulk system at unity filling using the experimental value of $J$ with vanishing dissipation are shown in Fig.~\ref{fig:3}\,(b), illustrating the revival dynamics of higher order correlations. 
In our spin system, we observe qualitatively similar dynamics, however with a lower amplitude due to the presence of dissipation (see Fig.~\ref{fig:3}\,(c)).
We find that $C^{(3)}_{1,-1}(t)$ evolves with a delay compared to $C^{(2)}_{1,-1}(t)$, in agreement with our calculation.
The build-up of multi-spin correlations by two-spin interactions can be understood because flipped spin pairs~\cite{Coherent_Zeiher} constrain the dynamics of neighboring spins during the Hamiltonian dynamics.


At later times $t_{\textrm{uv}}$, atom loss becomes dominant, which limits us from observing coherent revival dynamics.
From an independent experimental calibration, we extract the atom loss rate $\Gamma_{|\rightarrow\rangle}^{\textrm{ex}}=0.6(1)\,\si{\milli\second}^{-1}$, yielding a dressing quality factor of $J/\Gamma_{|\rightarrow\rangle}^{\textrm{ex}} \approx 2\pi \times 0.5$~\cite{SI}.
The observed value $\Gamma_{|\rightarrow\rangle}^{\textrm{ex}}$ is above the calculated value $\Gamma_{|\rightarrow\rangle}^{\textrm{th}} = 0.011\,\si{\milli\second}^{-1}$ assuming only off-resonant Rydberg and macrodimer scattering.
This additional loss could be associated with off-resonant excitation by the near-resonant sideband, depended on the detuning $\Delta$ and the power in the sideband, and was independent of the macrodimer coupling.
Possible origins include collective loss channels found in other Rydberg dressing experiments operating at high densities, potentially triggered by black-body radiation~\cite{Goldschmidt2015,Losses_Killian,Zeiher2016a}, as well as phase noise on the laser~\cite{festa2021blackbody}.

Besides atom loss, we identify a signature that is specific for macrodimers and their wave packets and limits dressing at low intermediate-state detunings.
Our spectroscopy of the lowest vibrational resonance starting from $\ket{\uparrow}$ revealed a surprisingly strong AC-Stark shift $V_{\textrm{pd}}$, see Fig.~\ref{fig:4}\,(a,b).
This originates from the coupling to a complete set of continuum modes for photodissociated states $\ket{\uparrow\! e},\ket{e\!\uparrow}$, see Fig.~\ref{fig:4}\,(c). 
Summing over the contributing modes and accounting for their kinetic energies $E_k = \hbar \omega_k$ in the relative motion and bound-continuum Franck-Condon factors~\cite{Photodiss_Deiglmayr}, we can predict the observed shift $V_{\textrm{pd}}$, see Fig.~\ref{fig:4}\,(b,d).
The shift increases for smaller carrier detunings $\Delta_{\textrm{C}} = \delta_0 - \Delta$ and adds an offset to the detunings used in the calculation of the spin coupling.
The coupling into the continuum furthermore broadens our resonance profiles, which will introduce dephasing to the spin dynamics at larger times $t_{\textrm{uv}}$, see the inset of~\ref{fig:4}\,(d).
We attribute this to the varying individual light shifts of the angular partial waves contributing to the oriented macrodimer as well as on-resonant photodissociation into continuum states for $ \Delta_{\textrm{C}} = -\omega_k$~\cite{SI}.
During our dressing experiment, we chose values $\Delta_{\textrm{C}}/2\pi = -6.3\,$MHz and $\omega_{\textrm{sb}} = 2\pi\times 726\,\si{\mega\hertz}$ where the effect of the broadening is small.

In conclusion, we realized Rydberg-dressed interactions restricted to a controllable selectable distance using macrodimers. 
At present, atom loss prevents us from observing coherent revivals. 
We anticipate an improvement by one order of magnitude in the dressing quality factor at unity Franck-Condon overlap.
This can be achieved using shallower binding potentials available at larger distances and principal quantum numbers~\cite{Barbier2021}.
Here, also motional states contribute less because the vibrational wave packets carry less kinetic energy.
In this scenario, we expect a preparation fidelity of $20\%$ for a cluster state in a system of 25 atoms.
In a cryogenic environment where losses approach the single-particle limit, this fidelity increases to $ 95\%$.
Further improvements include encoding the qubit in a clock state with larger Ramsey coherence time, increasing the power and reducing the noise on the UV laser, reducing the densities~\cite{Zeiher2016a,Coherent_Zeiher} or performing potential engineering~\cite{MW_control_Petrosyan}.  
Symmetrizing the spin couplings in the plane through magnetic field and polarization control promises the creation of large-scale two-dimensional cluster states.

\begin{acknowledgements}
\textbf{Acknowledgements:}
We thank all contributors to the open-source programs ``pair interaction'' and ``ARC''. Furthermore, we thank David Stephen for valuable discussions and Simon Evered for contributions to the experiment.
We acknowledge funding by the Max Planck Society (MPG) and from Deutsche Forschungsgemeinschaft (DFG, German Research Foundation) under Germany’s Excellence Strategy – EXC-2111 – 390814868 and Project No. BL 574/15-1 within SPP 1929 (GiRyd). 
This project has received funding from the European Union’s Horizon 2020 research and innovation programme under grant agreement No. 817482 (PASQuanS) and the European Research Council (ERC) No. 678580 (RyD-QMB).
K.S. acknowledges funding through a stipend from the International Max Planck Research School (IMPRS) for Quantum Science and Technology and J.R. acknowledges funding from the Max Planck Harvard Research Center for Quantum Optics. 
V.W. acknowledges support by the NSF through a grant for the Institute for Theoretical Atomic, Molecular,
and Optical Physics at Harvard University and the Smithsonian Astrophysical Observatory.
\end{acknowledgements}
\bibliography{Macrodimer_dressing_revtex}
\setcounter{figure}{0}
\renewcommand\theequation{S\arabic{equation}}
 \renewcommand\thefigure{S\arabic{figure}}
 \renewcommand\theHfigure{Supplement./thefigure}
\section{Supplementary information}
In this supplement, we discuss the electronic macrodimer states, the coupling from the molecular state into continuum modes, the calculation of the dressed interaction, the atom loss and further experimental details.

\subsection{Electronic macrodimer states}
We chose a $1_u$ macrodimer potential for Rydberg dressing.
The electronic part of the molecular wave function can be specified by the angular momentum projection along the molecular axis $\mathbf{R}$, which is $\pm 1$ for the chosen potential. 
The molecular bond length $R_\nu = 0.712(5)\,\si{\micro\m}\approx \sqrt{2}a_{\textrm{lat}}$ is close to the lattice diagonal distance, with $a_{\textrm{lat}} = 532\,\si{\nano \meter}$.
The small rotational constant $B_\ell = \hbar^2/(2\mu R_\nu^2) = h\times 229(4)\,\si{\Hz}$ with $\mu = m/2$ and $m$ the mass of a $^{87}$Rb atom justifies assuming that the interatomic orientation is conserved during the excitation.
The orientation of the coupled molecules relative to the magnetic field $\mathbf{B}$, which also points along one of the lattice diagonal directions, is either parallel or perpendicular.
Hence, all four possible molecular states can be labelled as $|\Psi^\nu_{\pm 1_{\parallel/\perp}}\rangle$, with $\nu$ the vibrational quantum number.

The molecular Rabi couplings $\Omega_\nu$ depend on the angle between the $\mathbf{B}-$field and the molecular frame and the light polarization, as discussed in Ref.~\cite{Structure_Tomography_Hollerith}. 
For $\mathbf{B}\parallel\mathbf{R}$, the initial two-atom ground state in the molecular frame is given by $\ket{\uparrow \uparrow} = |M_J=-1\rangle\otimes |M_I=-3\rangle$. 
Here, the electronic single-atom ground state is $\ket{\uparrow} = |5S_{1/2},F=2,m_F=-2\rangle$, with $M_J$ the total electronic angular momentum projection and $M_I$ the total nuclear spin projection.
In this configuration, the coupling to $|\Psi^{\nu}_{+1_\parallel}\rangle$ using two $\sigma^+$ photons reaches a maximum, while $|\Psi^{\nu}_{-1_\parallel}\rangle $ remains uncoupled.
For $\mathbf{B}\perp\mathbf{R}$, excitation rates to $|\Psi^{\nu}_{\pm 1_\perp}\rangle$ are finite but strongly suppressed.  
Hence, states $|\Psi^{\nu}_{\pm 1_\perp}\rangle$ and $|\Psi^{\nu}_{-1_\parallel}\rangle $ do not contribute to our dressing experiments.

For the optical coupling, we calculated a scaling factor $\alpha = 1.04$ to account for the difference in the electronic structure between the Rydberg state $|e\rangle = |36P_{1/2},m_J = 1/2\rangle $ and $|\Psi^{\nu}_{+1_\parallel}\rangle$.
Here, a small $R-$dependency of the electronic structure over the extension of the lowest vibrational modes was neglected. 
\begin{figure*}[htp]
\centering
  \includegraphics[width = 1 \textwidth]{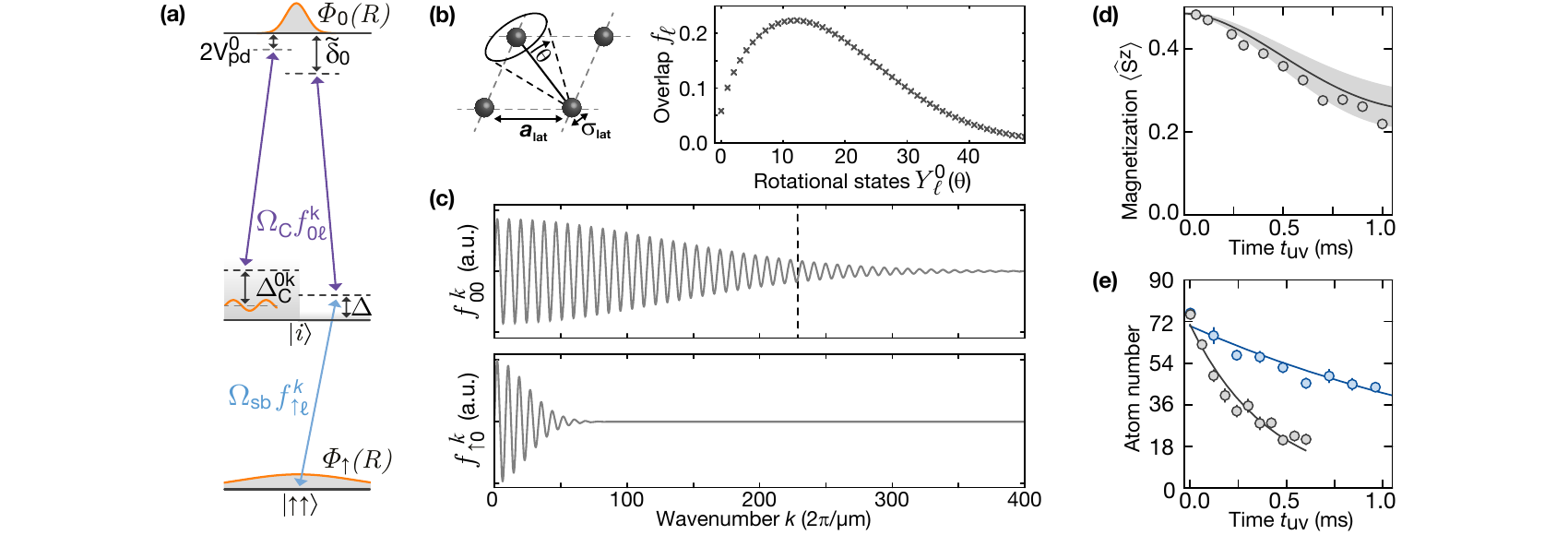}%
  \renewcommand\thefigure{S\arabic{figure}}
  \renewcommand\theHfigure{Supplement./thefigure}
  \caption{\label{fig:S1}  
Further details on the experiment.
\textbf{(a)} The coupling of the molecular state to the motional continuum energetically located above the bare electronic intermediate state $|i\rangle$ by the strong carrier field $\Omega_{\textrm{C}}$ causes an energy shift $V^0_{\textrm{pd}}$ (here for $\nu = 0$). 
On-resonant coupling into motional states at $\Delta^{0 k}_{\textrm{C}} = 0$ induces photodissociation.
In the two-photon excitation from the ground state $|\uparrow\uparrow\rangle$, only off-resonant continuum modes at low wavenumbers with finite overlap with $\Phi_\uparrow (R)$ as well as $\Phi_\nu (R)$ contribute. 
At two-photon detunings $\delta_0$, where the molecular state is only virtually populated, the ground state shifts by the spin-coupling $J^0_{\textrm{th}}$.
\textbf{(b)} The oriented relative wave packet in the lattice decomposes into angular momentum states $Y_\ell^0(\theta)$. \textbf{(c)} The overlap integral $f^k_{00}$ between the molecular wave packet $\Phi_0 (R)$ and the continuum modes (here for $\nu = \ell = 0$) shows contributions at much higher wavenumbers $k$ where the corresponding integral $f_{\uparrow 0}^k$ vanishes. 
The dashed black line indicates the $k-$value where $\Delta^{0 k}_{\textrm{C}} = 0$ for our dressing parameters. 
While the $\ell -$independent envelope of $f_{\nu\ell}^k$ reproduces the shift $V^0_{\textrm{pd}}$, the $\ell -$dependent oscillations contribute to the broadening. 
\textbf{(d)} Magnetization dynamics for the same dataset as Fig.~\ref{fig:2} and Fig.~\ref{fig:3}. Again, the solid line was calculated using the parameters obtained from fitting $C^{(2)}_{1,-1}(t)$ and the shaded area is the theoretical expectation. \textbf{(e)} Decay calibration curves for all atoms in $\ket{ \uparrow }$ (gray) as well as all atoms in the relevant initial state $\ket{\rightarrow } $ used in the dressing sequences (blue). From the fit, we extract 
$\Gamma^{\textrm{ex}}_{|\rightarrow\rangle} = 0.6(1)\,\si{\milli \second}^{-1}$.
}\end{figure*}

\subsection{Motional states}
The total ground state $|\Psi_{\uparrow \uparrow}\rangle$ consists of a electronic part $\ket{\uparrow \uparrow }$ and a relative wave packet $\Phi_\uparrow(\mathbf{R})$ provided by the motional ground state in the optical lattice, at a distance of a lattice diagonal.
The lattice depth was $1000\,E_\textrm{r}$ for all three directions, with $E_\textrm{r} = h^2/(8ma^2_{\textrm{lat}})$ the recoil energy. 
The small single-particle width $\sigma_{\textrm{lat}} = 21\,\si{\nano \meter} \ll \sqrt{2}a_{\textrm{lat}}$ allows us to separate $\Phi_\uparrow (\mathbf{R})\approx \Phi_\uparrow (R)\psi_\uparrow (\theta)$, with the radial part $\Phi_\uparrow (R)\propto e^{-(R-\sqrt{2}a_{\textrm{lat}})^2/(8\sigma_{\textrm{lat}}^2)}$ and the angular part $\psi_\uparrow (\theta) \propto e^{-\left(\sin(\theta/2)a_{\textrm{lat}}/\sigma_{\textrm{lat}}\right)^2}$, see Fig.~\ref{fig:S1}\,(b).
Because all lattice depths were equal, the trapping was isotropic and the angular distribution only depended on the angle $\theta$.
We decompose $\psi_\uparrow (\theta)=\sum_{\ell} f_\ell Y^0_\ell (\theta)$ into spherical harmonics $Y^0_{\ell} (\theta)$ and find contributions for $\ell = 0,1,...,\leq 50$ and zero angular momentum projection.

In the molecular state $|\Psi^{\nu}_{+1_\parallel}\rangle$, interatomic motion enters via the narrow vibrational state $\Phi_\nu (R)$ and the rotational states $Y^0_\ell (\theta)$. 
We expand $|\Psi^{\nu}_{+1_\parallel}\rangle = \sum_\ell f_\ell |\Psi^{\nu\ell}_{+1_\parallel}\rangle$, with $|\Psi^{\nu\ell}_{+1_\parallel}\rangle$ the contribution of the rotational angular momentum $\ell$ to the aligned molecular state.
The width of the wave packet is $\sigma_0 = 5.5\,\si{\nano\m}$, yielding a normalized wave packet $\Phi_{0} (R) \approx 1/\left(2\pi R^4_0\sigma_0^2\right)^{\frac{1}{4}}e^{-\left(R-R_\nu\right)^2/4\sigma_0^2}$.

After photodissociation, the non-interacting continuum eigenstates are $\psi^k_\ell(\mathbf{R}) = j_\ell(kR)Y^0_\ell(\theta)$, with $k$ the wavevector in the interatomic motion and $j_\ell(kR)$ the spherical Bessel functions.
We found that the coupling to the motional continuum was independent of the underlying optical lattice and assumed kinetic energies $E_k = \hbar^2k^2/(2\mu) = \hbar \omega_k$.
A decomposition of the ground state $\Phi_\uparrow (\mathbf{R})$ into continuum modes yields Franck-Condon factors
\begin{align}
 f^{k}_{\uparrow\ell}  = f_\ell \int_0^\infty j_\ell(kR) \Phi_\uparrow(R) R^2dR.
\end{align}
Similarly, a decomposition of the molecular state $\Phi_\nu (R)$ yields
\begin{align}
 f^{k}_{\nu\ell}  = \int_0^\infty j_\ell(kR) \Phi_\nu(R) R^2dR,
\end{align}
which both consist of a rapidly oscillating term and a broad envelope $ \propto e^{-2k^2\sigma_0^2}$, see Fig.~\ref{fig:S1}\,(c).

\subsection{Excitation model}
In the two-photon excitation scheme, using $\sigma^+-$polarized light, only the single-particle Rydberg state $|e \rangle$ contributes to the two electronic intermediate states $\ket{\uparrow\!\!\!e}$ and $\ket{e\!\!\!\uparrow }$, which are combined into $|i\rangle = 1/\sqrt{2}\left(|e\!\uparrow\rangle + |\!\!\uparrow \!e\rangle\right)$. 
The antisymmetric state remains uncoupled. 
The total state of the system can be described by
\begin{align}
|\Psi_{\textrm{tot}}(t)\rangle =~& c_{\uparrow\uparrow}(t)|\Psi_{\uparrow\uparrow}\rangle  + \sum_{\ell k}~ c_{k\ell}(t)|i^k_\ell\rangle \notag\\+& \sum_{\nu\ell} c_{\nu\ell}(t) |\Psi^{\nu\ell}_{+1_\parallel}\rangle,
\end{align}
where $k$ and $\ell$ specify the previously introduced motional states.

The coupling scheme relies on modulating a sideband at a frequency $\omega_{\textrm{sb}}$ and relative field amplitude $\beta$ on our carrier field using an electro-optical modulator (EOM).
The ground state is then coupled to the intermediate state by $\Omega_{\textrm{sb}} = \sqrt{2}\beta\Omega $, which is in turn coupled to the molecular state by $\Omega_{\textrm{C}} = \sqrt{2}\alpha\Omega $, see Fig.~\ref{fig:S1}\,(a).
Here, $\Omega$ is the experimentally calibrated single-atom Rabi frequency between $\ket{\uparrow}$ and $|e\rangle $, coupled by the carrier field.
Assuming real-valued Rabi frequencies, the Hamiltonian of the system in a rotating frame writes
\begin{align}
\label{eq:Hamil}
& \hat{H} = \frac{\Omega_{\textrm{sb}}}{2}\sum_{\ell k}\left( f^{k}_{\uparrow \ell} |\Psi_{\uparrow\uparrow}\rangle \langle i^k_\ell| + h.c. \right) - \sum_{\ell k} \Delta_k |i^k_\ell\rangle \langle i^k_\ell |  \\ & +  \frac{\Omega_{\textrm{C}}}{2}\sum_{\nu \ell k}\left( f^{k}_{\nu \ell} |\Psi^{\nu\ell}_{+1_\parallel}\rangle \langle i^k_\ell| + h.c. \right) - \sum_{\nu\ell} \delta_\nu |\Psi^{\nu\ell}_{+1_\parallel} \rangle \langle \Psi^{\nu\ell}_{+1_\parallel}|. \notag
\end{align}
Here, the overlap integrals $ f^{k}_{\uparrow\ell}$ and $ f^{k}_{\nu\ell}$ account for the motional states and  $\Delta_k = \Delta - \omega_k$ is the intermediate state detuning including the energy of the continuum modes.
Furthermore, $\delta_\nu = \delta_0-\nu \hbar \omega_{\nu}$ is the two-photon detuning to different vibrational states, which have an energy spacing $\omega_{\nu}/2\pi = 3.80\,$MHz that is almost independent of $\nu$.

\subsection{Elimination of the continuum}
The large carrier Rabi frequency $\Omega_{\textrm{C}}$ and a complete set of available continuum states induces a non-negligible energy shift on the vibrational resonances, especially for $\nu = 0$. 
Hence, we dress the macrodimer states with the continuum first and then calculate the molecular Rabi coupling $\Omega_\nu$ and the spin coupling $J_{\textrm{th}}$.

For an initially prepared macrodimer with $c_{\nu\ell}(0) = 1$ at a fixed rotational quantum number $\ell$ and a single-photon carrier detuning $\Delta_{\textrm{C}}^{\nu k} = \Delta^{\nu}_{\textrm{C}} + \omega_k $ with $\delta_\nu = \Delta + \Delta^\nu_{\textrm{C}}$, the time-dependent Schrödinger equation yields
\begin{align}\label{eq:adel}
i\frac{d}{dt}\,c_{k\ell}(t)&= \Delta_{\textrm{C}}^{\nu k}\, c_{k\ell}(t) + \frac{\Omega_{\textrm{C}}}{2} c_{\nu\ell}(t) f^{k}_{\nu\ell} \notag\\
i\frac{d}{dt}\,c_{\nu\ell}(t)  &=   \frac{\Omega_{\textrm{C}}}{2}  \,\sum_k c_{k\ell}(t) f^{k}_{\nu\ell},
\end{align} 
with a dominant off-resonant part $\Delta_{\textrm{C}}^{\nu k} \gg  \Omega_{\textrm{C}}f^{k}_{\nu\ell}$ and a small on-resonant part at $\Delta_{\textrm{C}}^{\nu k} = 0$.

Assuming a stationary population $\frac{d}{dt}\,c_{k\ell}(t) = 0$ and neglecting carrier-mediated two-photon couplings between different vibrational states, the above equations become
\begin{align}\label{shift_brd}
i\frac{d}{dt}c_{\nu\ell} = \int_0^\infty dk \frac{\rho |\Omega_{\textrm{C}}f^{k}_{\nu\ell}|^2}{4\Delta^{\nu k}_{\textrm{C}}} c_{\nu\ell} = 2(V^{\nu\ell}_{\textrm{pd}}  + i\gamma^{\nu\ell}_{\textrm{pd}})c_{\nu\ell},
\end{align}
where a sufficiently large system size allows to replace $\sum_k \rightarrow \int \rho \,dk$, with $\rho$ the density of states along the radial coordinate.
We calculate the complex valued integral by exploiting the Sokhotski-Plemelj theorem, where the real part $V^{\nu\ell}_{\textrm{pd}}$ shifts the energy of the dressed macrodimer states, see Fig.~\ref{fig:4} and Fig.~\ref{fig:S1}\,(a).
This modifies the two-photon detunings via $\widetilde\delta^\ell_\nu = \delta_\nu - 2V^{\nu\ell}_{\textrm{pd}}$.
For $\nu=0$, the imaginary part yields
\begin{align}\label{mot_brd}
2\gamma^{0\ell}_{\textrm{pd}} = \frac{\Omega_{\textrm{C}}^2\sigma_0}{2}\sqrt{\frac{4\pi\mu}{\hbar\Delta^0_{\textrm{C}}}}e^{-4\mu|\Delta^0_{\textrm{C}}|\sigma_0^2/\hbar}.
\end{align}
This represents an on-resonant photodissociation rate and is identical to the result obtained from Fermi's Golden rule. 
The $\ell -$dependency of $V^{\nu\ell}_{\textrm{pd}}$ and $\gamma^{\nu\ell}_{\textrm{pd}}$ arises from the oscillatory behavior of $f^{k}_{\nu\ell}$ at wavenumbers close to the divergence in Eq.~(\ref{shift_brd}), see Fig.~\ref{fig:S1}\,(c).

\subsection{Resonance profile}
Here, we calculate excitation rates into an isolated continuum-dressed macrodimer state for an initially prepared ground state atom pair at small two-photon detunings $\delta^\ell_\nu$. 
The dynamics is described by
\begin{align}
\label{eq:supthreeeq}
i\frac{d}{dt}\,c_{\uparrow\uparrow}(t)  &=   \frac{\Omega_{\textrm{sb}}}{2} \sum_{\ell k} f^{k}_{\uparrow\ell}\, c_{k\ell} (t)\notag\\
i\frac{d}{dt}\, c_{k\ell}(t) &= -\Delta_k  \, c_{k\ell}(t) +\frac{\Omega_{\textrm{C}}}{2} f^{k}_{\nu\ell} \, c_{\nu\ell}(t) \notag \\ &   +   \frac{\Omega_{\textrm{sb}}}{2} f^{k}_{\uparrow\ell} \,c_{\uparrow\uparrow} (t) \\
i\frac{d}{dt}\,c_{\nu\ell}(t)&= \left(-\delta^\ell_\nu -i\gamma_{\textrm{lp}}  \right)\, c_{\nu\ell}(t)\notag \\ & + \frac{\Omega_{\textrm{C}}}{2} \sum_k f^{k}_{\nu\ell}\, c_{k\ell}(t). \notag
\end{align} 
Here, $2\gamma_{\textrm{lp}} = 2\pi \times 920\,$kHz is the experimental linewidth at low power, which is limited by the laser, the lattice inhomogeneity, the rotational states and Doppler broadening from the vibration.

Close to the two-photon resonance, the hierarchy of energy scales $|\Delta| \gg \delta_\nu,f_{\nu\ell}^k|\Omega_{\textrm{C}}|,f_{\uparrow\ell}^k|\Omega_{\textrm{sb}}|$ allows to adiabatically eliminate the intermediate state in Eq.~(\ref{eq:supthreeeq}). In analogy to the previous section, we assume $i\frac{d}{dt}\, c_{k\ell}(t) = 0$ and insert the obtained expression for $c_{k\ell}(t)$ in the remaining two equations, which again provides $V^{\nu\ell}_{\textrm{pd}}$ and $\gamma^{\nu\ell}_{\textrm{pd}}$.
We furthermore neglect a term quadratic in $\Omega_{\textrm{sb}}$ which represents a small energy shift in the ground state and get
\begin{align}\label{eq:res_prof}
i\hbar\frac{d}{dt}\,c_{\uparrow\uparrow}(t)  &=   \sum_{\ell}\, \frac{\Omega^{\ell}_{\nu}}{2}  \, c_{\nu\ell}(t) \notag\\
i\hbar\frac{d}{dt}\,c_{\nu\ell}(t)&= \left[-\widetilde\delta^\ell_\nu -i(\gamma_{\textrm{lp}} + \gamma^{\nu\ell}_\mathrm{pd}) \right]\, c_{\nu\ell}(t)\notag\\
&~ + \frac{\Omega^{\ell}_{\nu}}{2} c_{\uparrow\uparrow} (t),
\end{align} 
where $\Omega^{\ell}_{\nu}$ is the effective two-photon Rabi frequency for $|\Psi^{\nu\ell}_{+1_\parallel}\rangle$.
For oriented molecular states $|\Psi^{\nu}_{+1_\parallel}\rangle$, the coupling rate $\Omega_\nu = \sum f_\ell \Omega^\ell_\nu$ is
\begin{align}\label{two_phot_r}
\Omega_{\nu} = \Omega_{\textrm{sb}}\Omega_{\textrm{C}} \sum_{\ell k} f_\ell\frac{f^{k}_{\uparrow\ell} f^{k}_{\nu\ell}}{2\Delta_k} \approx \frac{\Omega_{\textrm{sb}}\Omega_{\textrm{C}}}{2\Delta} f_\uparrow^\nu .
\end{align}
The small kinetic energy in the ground state restricts contributions from $f^{k}_{\uparrow\ell}$ to small wavenumbers where $\Delta_k \approx \Delta$ is effectively $k-$independent and the above sums become $ \sum_{k} f^{k}_{\ell\uparrow} f^{k}_{\nu\ell} = f_\ell f^\nu_\uparrow = f_\ell \int_0^\infty \Phi_\uparrow(R) \Phi_\nu(R) R^2dR$ and $\sum_{\ell} |f_\ell|^2 = 1$, see Fig.~\ref{fig:S1}\,(b)). 
Hence, continuum states play a minor role in the calculation of $\Omega_{\nu}$. 

A steady-state analysis of the total macrodimer population $\sum_\ell |c_{\nu\ell}|^2$ asuming $c_{\uparrow \uparrow}(t)\approx 1$ yields a resonance profile
\begin{align}\label{brd}
\Gamma^{\nu}(\delta) \propto \sum_\ell\frac{|f_\ell|^2}{\left(\frac{\delta_\nu}{2} - V^{\nu\ell}_{\mathrm{pd}}\right)^2+\frac{1}{4}\left(\gamma_{\textrm{lp}}+\gamma^{\nu\ell}_{\textrm{pd}}\right)^2}.
\end{align}
The observed shift is $V^\nu_{\mathrm{pd}} \approx \sum_\ell |f_\ell|^2 V^{\nu\ell}_{\mathrm{pd}}$ and broadening emerges due to $\ell -$dependent resonances, as indicated by the blue shaded area in Fig.~\ref{fig:4}\,(d).
Additionally, $\gamma^\nu_{\textrm{pd}}  \approx \sum_\ell |f_\ell|^2 \gamma^{\nu\ell}_{\textrm{pd}}$ directly increases the width of the individual resonances.

Note the factor of 2 appearing when replacing $\delta_\nu$ by our experimental value for $\Delta$.
In our two-photon spectroscopy shown in Fig.~\ref{fig:4}, we tune the overall frequency of the UV laser and keep the modulation frequency $\omega_{\textrm{sb}}$ constant. 
Hence, changing $\Delta$ in our experiments results in a change in $\delta_\nu$ which is twice as high.
\subsection{Dressed interactions}
Here, we calculate the spin coupling $J_{\textrm{th}}$ for an atom pair $|\!\!\uparrow\uparrow\rangle$ for off-resonant coupling $\Omega_\nu \ll \widetilde \delta_\nu $ and $ \Omega_{\textrm{sb}} \ll \Delta$.
The contributions of the vibrational states to the spin coupling are
\begin{align}\label{eq:fourth_order}
J_{\textrm{th}}^\nu = \sum_{\ell k k^\prime} \frac{\Omega^2_{\textrm{C}}\Omega^2_{\textrm{sb}} f^k_{\uparrow \ell} f_{\nu\ell}^k f^{k^\prime}_{\nu \ell} f_{\uparrow \ell }^{k^\prime}  }{16\widetilde\delta^\ell_\nu\Delta_k\Delta_{k^\prime}} \approx  \frac{\Omega_\nu^2}{4\widetilde\delta_\nu}.
\end{align}
The left expression is obtained by applying fourth order perturbation theory in the coupling parameter. 
The approximation on the right side assumes $k-$independent detunings where sums over $k$ and $k^\prime$ collapse to identities due to the Franck-Condon integrals $f^{k}_{\uparrow\ell}$, as discussed in the context of Eq.~(\ref{two_phot_r}).
It furthermore neglects the $\ell-$dependence of the shift and is correct up to $95\%$ for our chosen parameters.

During the dressing sequence, our parameters were $\Omega/2\pi = 2.83(5)\,$MHz, $\omega_{\textrm{sb}}/2\pi = 726\,$MHz and $\beta = 0.062$.  
The interaction shift of the lowest vibrational state relative to the state $|ee\rangle$ is $U^0 = 2\pi \times 735.3(1)\,$MHz, see Fig.~\ref{fig:1}\,(b).
The Ising interaction term critically depends on $\Delta$ and  $\widetilde\delta_\nu$.
They were calibrated using the continuously tracked position of the lowest macrodimer resonance and the working point of the laser during dressing, as well as the shifts $V^0_{\textrm{pd}}$ at both laser frequencies.
This provides $\Delta = 2\pi \times 3.58\,$MHz and $\widetilde\delta_0 = -2\pi \times 3.01\,$MHz.
The total spin interaction is $J_{\textrm{th}} = \sum_\nu J_{\textrm{th}}^\nu =  -2\pi \times 370(40)\,$Hz, which is dominated by $J_{\textrm{th}}^0 \approx 2/3 J_{\textrm{th}}$ due to the small detuning and the large Franck-Condon overlap.
The calibration error on the calculated value $J_{\textrm{th}}\propto \Omega^4$ was dominated by uncertainties in the Rabi frequency $\Omega$.
Other contributions include an uncertainty in the sideband power, small temporal drifts of our macrodimer resonance, the electronic structure of the molecular state, the bond length $R_\nu$ and the lattice depth.

The soft-core interaction $J_{\textrm{sc}}$ shown in Fig.~\ref{fig:1}\,(a) was calculated using reasonable values for conventional Rydberg dressing for our principal quantum number and a Rabi frequency and detuning in the weak dressing regime.
During macrodimer dressing, conventional long-range Rydberg dressing potentials induced by the weak sideband or the far-off-resonant carrier were below $5\,$Hz and neglected.

\subsection{Atom loss}
Assuming only off-resonant scattering, the theoretical loss rate per atom $\Gamma_{|\rightarrow\rangle}^{\textrm{th}}$ was estimated by
$\Gamma_{|\rightarrow\rangle}^{\textrm{th}} \approx \frac{\gamma_{\textrm{e}}}{2}\left(\frac{\beta\Omega}{2\Delta}\right)^2 + \frac{\gamma_{\nu}}{2}\sum_{\nu}\left(\frac{\Omega_\nu}{2\delta_\nu}\right)^2$,
with $\gamma_{\textrm{e}} = 23\,\si{\milli \second}^{-1}$ and $\gamma_{\nu} \approx 2\gamma_{\textrm{e}}$~\cite{Overstreet2009} the radiative decay rates of the Rydberg state and the macrodimer state.
Factors of $1/2$ account for the probability to be in the coupled state $\ket{\uparrow }$ and the fact that macrodimer excitation removes two atoms. 
The calculated $\Gamma_{|\rightarrow\rangle}^{\textrm{th}} = 0.011\,\si{\milli\second}^{-1}$ during our dressing sequence is dominated by single-atom loss.
Experimentally, we find a higher loss rate $\Gamma^{\textrm{ex}}_{|\rightarrow\rangle} = 0.6(1)\,\si{\milli \second}^{-1}$ by fitting an exponential decay $N(t)=N_0 e^{-\Gamma_{|\rightarrow\rangle}^{\textrm{ex}} t}$ to the atom loss in a reference measurement, see Fig.~\ref{fig:S1}\,(d).  
We could relate the additional loss to the near-resonant sideband.
We do not observe correlated pair loss during dressing, even though photodissociation via $\gamma^0_{\textrm{pd}}$ is expected to increase the associated loss rate. 

\subsection{Calculated spin dynamics}
The calculation of $C^{(2)}_{1,-1}(t)$ and $C^{(3)}_{1,-1}(t)$ including dissipation was performed with QuTip~\cite{QuTip}. 
We accounted for atom loss by projecting dressed atoms $\ket{\rightarrow }$ into a state $|0\rangle $ representing a lost atom at a rate $\Gamma^{\textrm{ex}}_{|\rightarrow\rangle}$ ~\cite{Zeiher2016a}. 
Our spin-resolved detection relies on removing all atoms in $\ket{ \uparrow }$ and then imaging the atoms in $\ket{\downarrow}$.
Hence, lost atoms are identified as $\ket{\uparrow }$.
In the model, we only account for uncorrelated atom loss. 
We did not account for UV-mediated projections from $\ket{ \uparrow }$ to $\ket{ \downarrow }$ due to the absence of retrapping~\cite{Coherent_Zeiher}.
At $t_{\textrm{uv}}=0$, we initialize $94\% $ of the atoms in $|0 \rangle $ to account for the lattice filling.
The results are shown in Fig.~\ref{fig:2} and Fig.~\ref{fig:3}.
Fitting the observed correlations $C^{(2)}_{1,-1}(t)$ to the same model, we obtain $J = 2\pi \times 318(20)\,$Hz and $\Gamma^{\textrm{fit}}_{|\rightarrow\rangle} = 0.46(5)\,\si{\milli \second}^{-1}$ for the two free parameters. 
Errors on the parameters $J$ and $\Gamma^{\textrm{fit}}_{|\rightarrow\rangle}$ were estimated by the $1\sigma-67\%$ confidence interval of the fit.
The obtained value $\Gamma^{\textrm{fit}}_{|\rightarrow\rangle}$ is close to the measured value $\Gamma^{\textrm{ex}}_{|\rightarrow\rangle}$ but above $\Gamma^{\textrm{th}}_{|\rightarrow\rangle}$.
We also calculate the magnetization $\langle \hat{S}^z(t)\rangle = \frac{\langle N(t) \rangle - N_0/2}{N_0}$, with $N_0$ the initial atom number and $\langle N(t) \rangle$ the averaged atom number after the Ramsey sequence, see Fig.~\ref{fig:S1}\,(d). 
Again, the calculated dynamics agrees with the calculation. 
The calculated spin dynamics was evaluated in the bulk, which evolves twice as fast as spins at the edge~\cite{Coherent_Zeiher}. 
Our observed correlations were calculated in a centered region of $11\,\textrm{x}\,11$ lattice sites.

$•$
\subsection*{Three-spin correlation function}
The spatially averaged connected three-spin correlator is given by 
\begin{align}
\begin{split}
& C^{(3)}_{\mathbf{R}_1\mathbf{R}_2} = \biggl ( \biggl\langle \biggl (\hat{S}^z_{\mathbf{R}^\prime}-\langle\hat{S}^z_{\mathbf{R}^\prime}\rangle \biggr)
\biggl(\hat{S}^z_{\mathbf{R}^\prime+\mathbf{R}_{1}} - \\ &\langle\hat{S}^z_{\mathbf{R}^\prime+\mathbf{R}_{1}}\rangle\biggr)  \biggl(\hat{S}^z_{\mathbf{R}^\prime + \mathbf{R}_{2}}- \langle\hat{S}^z_{\mathbf{R}^\prime+\mathbf{R}_{2}}\rangle\biggr) \biggr\rangle \biggr ) _{ \mathbf{R}^\prime}\end{split}
\end{align}
where $(.)_{\mathbf{R}}$ denotes spatial averaging and $\langle . \rangle$ averaging over experimental shots.
\end{document}